\documentclass[fleqn,12pt,twoside]{article}
\usepackage[headings]{espcrc1}
\usepackage{epsfig}

\readRCS
$Id: espcrc1.tex,v 1.2 2004/02/24 11:22:11 spepping Exp $
\usepackage{graphicx}

\newcommand{\AmS}{{\protect\the\textfont2
  A\kern-.1667em\lower.5ex\hbox{M}\kern-.125emS}}

\title{Focus talk on interactions between jets and medium}

\author{J\"org Ruppert
\address{Department of Physics, Duke University, Durham, NC 27708, USA}
}
       
\begin{document}

\maketitle

\begin{abstract} 
The energy and momentum lost by a hard parton propagating through hot and dense matter has to be redistributed during the nuclear medium evolution. Apart from heating the medium, there is the possibility that collective modes are excited leading to the emergence of Mach cones or Cherenkov radiation. Recent two-particle correlation measurements by STAR \cite{Wang:2004kf} and PHENIX \cite{Adler:2005ee} at RHIC indicate that such phenomena may play an important role in understanding the jet-medium interactions.
Possible collective modes are discussed and it is demonstrated that Mach cones as created by colorless or colored sound are a possible explanation of the hardronic two-particle correlation data.
\end{abstract}

\section{Introduction}

QCD jet quenching in relativistic heavy ion collision was proposed as a promising tool \cite{Wang:1991xy,Pluemer:1995dp} to study the properties of a new state of matter - distinct from ordinary hadronic matter - expected to be created in ultrarelativistic heavy-ion collisions at RHIC. 
The energy loss of hard partons created in the first moments of the collision has therefore been studied extensively, see e.g. \cite{Accardi:2003gp}. However in order to fully exploit those tools, the influence of the medium evolution on the energy loss has to be understood in a more quantitative way, see e.g. \cite{Renk:2005ta} .

Recently, measurements of two-particle correlations involving one hard trigger particle 
have shown a surprising splitting of the away side peak for all centralities but peripheral collisions, qualitatively very different from a broadened away side peak as observed in p-p or d-Au collisions, see e.g. \cite{Wang:2004kf,Adler:2005ee}. In proceeding contribution I focus on theoretical ideas how those results could be interpreted in terms of the redistribution of energy and momentum to the medium during its dynamical evolution. 

Pal and Pratt pointed out in \cite{Pal:2003zf} that the suppressed
away-side jet's energy and momentum survives during the medium evolution. They argued that the fate of away-side jets could be studied by analyzing the momentum distributions of soft hadrons relative to the jet axis. Furthermore they argued that the actual mechanism of this energy and momentum redistribution might be due to interaction of the radiated gluons that are formed at space-time points along the parent jet's trajectory with the evolving medium. However, the statement that $100\%$ of the energy lost by the away-side jet is thus eventually transformed into thermal energy, while only maintaining the momentum transferred to the medium, faces a challenge by the away-side peak structure 
observed in two-particle correlations from the PHENIX \cite{Adler:2005ee} and STAR collaboration \cite{Wang:2004kf}, see \cite{Renk:2005si}.

Because of those experimental results the question of what part of the energy lost by a jet is simply 'heating' the plasma and what part is redistributed into (possible) collective excitations of the medium has aroused growing interest in our scientific community. 

Some of those collective excitations can provide a possible explanation of the structures in the two-particle correlation data as will be discussed below.


\section{Collective excitations of the medium}

One can subsume the proposed collective excitations of the nuclear medium under two sub-classes, namely colorless (hydrodynamical) and colorful modes. 
St\"ocker argued in \cite{Stoecker:2004qu} (without using this nomenclature) that both modes might play an important role in understanding two-particle correlation measurements of secondary particles in RHIC experiments. 

Colorless hydrodynamical excitation in the sQGP leading to what could be called 'conical flow' where studied in linearized hydrondynamics by Casalderrey-Solana, Shuryak, and Teaney, for details see \cite{Casalderrey-Solana:2004qm,Casalderrey-Solana:2005QM}.  The idea is that a substantial part of the radiative energy could be deposited in a collective sound mode and that the interference of perturbations from a supersonically moving partonic jet creates conical flow behind it in the form of shock-waves leading to Mach cone density structures in the medium.

Also the possibility of colorful excitations in the plasma medium has been studied recently, namely longitudinal and transverse modes. Those collective excitations of the medium can theoretically associated with effective longitudinal and transverse gluon dispersion relations. If those extend into the space-like region, these dispersion relations can account for the emergence of Mach cones in the wake of the jet or Cherenkov (like) gluon radiation, respectively. Mach cones are associated with collective motion of part of the medium whereas Cherenkov (like) gluon radiation is emerging as gluonic shower.

The longitudinal modes could be called 'colored sound' and allow - as discussed above in the case of colorless sound - for the formation of Mach cone structures in their wake if the leading parton is travelling supersonically in the medium. B. M\"uller and I studied that
possibility by considering the plasma's reaction to a charged colored particle traversing 
the medium by calculating the plasma's response to that particle. We calculated the properties of the current and charge density wake that is induced by such a partonic jet \cite{Ruppert:2005uz}. We used methods of linear response theory and assumed the medium to be isotropic and homogeneous. In such a framework non-abelian effects are included indirectly via the dielectric functions, $\epsilon_L$ and $\epsilon_T$. Linear response theory implies that the dielectric functions are not in turn modified by the effects of the external current.

Two qualitatively different scenarios were investigated: in the first it is assumed that the plasmon is in a high temperature regime where the gluon self-energy can be described using a high temperature expansion \cite{Weldon:1982aq} $T \gg \omega, k$. This is equivalent to the HTL approximation and the dielectric functions are therefore gauge invariant \cite{Pisarski:1989cs}.
They lead via dispersion relations to longitudinal and transverse plama modes that can only appear in the time-like sector of the $\omega, k$-plane. 
On this scenario the wake induced by a partonic jet reveals a charge and current density profile in which no Mach cones appear, but the charge carries a screening color cloud along with it. Fig. 1 in \cite{Ruppert:2005uz}  shows the charge density of a colored parton traveling with $v=0.99~c$ in cylindrical coordinates. 

In the other scenario we assume that the plasma is in a regime where it exhibits properties of a quantum liquid. Since there is a lack of methods to calculate the dieletric function in such a regime, we use a simple model. Nonetheless, this simplified model is constructed in such a way that it allows for a general conclusion quite independent from the exact form of the dielectric functions. To be specific it is assumed that the longitudinal plasmon of the quantum liquid like plasma has the following Bloch \cite{Bloch} dispersion relation  $\omega=\sqrt{c_s^2 k^2 + \omega_p^2}$, where $\omega_p$ denotes the plasma frequency and $c_s$ the speed of colored sound in units of $c$. It is assumed that a critical momentum $k_c$ exists which separates the regimes of collective and single particle modes in the quantum liquid, therefore this collective mode exist for $k\le k_c$ only.
Such a plasmon dispersion relation is realized by the following longitudinal dielectric function:
$\epsilon_L=1+\frac{\omega_p^2/2}{c_s^2k^2-\omega^2+\omega_p^2/2}\,\,(k\le k_c)\,$.
Note that this differs from the classical, hydodynamical function of Bloch since 
the latter is singular at small $k$ and $\omega$.
The Bloch (like) plasmon mode extends for $k>\omega_p/\sqrt{1-c_s^2}$ into the space-like region of the $\omega,k$ plane. Notice that this is different from the high-temperature plasma, where longitudinal and transverse plasma modes only appear in the time-like region $|\omega/k|>1$. In the quantum liquid scenario one can expect that the modes with low phase velocity $|\omega/k|<c_s$ suffer severe Landau damping because they accelerate he slower moving charges and decelerate those moving faster that the wave. A charge moving with a velocity that is lower than the speed of colored sound can only excite these modes and not the modes with intermediate phase velocities $c_s<|\omega/k|<1$, which are undamped. The qualitative properties of the color wake can in the case of a subsonically traveling jet expected to be analogous to those of the high temperature plasma, namely that the charge carries only a screening color cloud with it and Mach cones do not appear. If the jet is traveling supersonically the situation changes and the modes with intermediate phase velocities can be excited. Mach cones can appear. 
To illustrate this we consider a speed of colored sound of $c_s=1/\sqrt{3}$ and calculated the charge density for a colored particle traveling with $v/c=0.99>c_s$, see Fig. 3 in \cite{Ruppert:2005uz}.

We now turn to the idea of transverse colorful excitations leading to Cherenkov radiation in the medium induced by a superluminally traveling jet. This idea was proposed by Dremin \cite{Dremin:1979yg} and has been investigated also by Majumder and Wang \cite{Majumder:2005sw}.
They considered gluon bremsstrahlung induced by multiple parton scattering in a finite dense medium. They showed that if the transverse dielectric constants leads to a {\it space-like} transverse dispersion the LPM interference pattern becomes Cherenkov-like with an increased opening angle determined by the dielectric  constant. One can find the cone angle of the Cherenkov-like gluon radiation to correspond exactly to the angle of classical Cherenkov radiation in the soft radiation limit $z \sim 0$, namely $\cos^2\Theta_c=1/\epsilon_L$.  The condition that the transverse mode has to extend into the space-like region is necessary for the existence of Cherenkov-like radiation.

Koch, Majumder, and Wang argued in \cite{Koch:2005sx} that the possibility of the occurrence of Cherenkov-like radiation is sensitive to the presence of partonic bound states and illustrate in a model calculation how the dispersion relation of a massless particle coupled to two different massive resonance states changes. They also discuss the dependence of the Cherenkov angle on the momentum of the emitted particles, see Fig. 5 in \cite{Koch:2005sx}. Note that the momentum scale on the axis on this figure is in units of $ T$. Therefore emitted particles with momenta with above $\approx 100 {\rm MeV}$ lead to a peak in the Cherenkov-like emission pattern at angles $\Theta_c$ relative to the leading parton of {\it less} than $15$ degrees. The change of that angle with the momentum of the emitted particle is also very low in that range. Although it is not clear how this emission pattern of the Cherenkov-like gluon shower would survive hadronization, considering the peak structure in the two-particle correlations  around $90 \pm \gg 50$ degrees this scenario faces a hard experimental challenge \cite{Adler:2005ee,Wang:2004kf} if one would infer it to explain those structures. 
Although such a colorful transverse collective mode and therefore Cherenkov-like gluon radiation could appear, it seems unlikely to be the primary mechanism explaining hadronic correlation measurements. 

An interesting question is if (and if possible how) colorful modes could eventually leave their marks during the evolution of the medium when it hadronizes and is already hadronized. One could e.g. expect that a colored sound leading to a Mach cone charge density wake could eventually transfer in part energy and momentum to colorless sound in the hadronic phase. Those questions have to be addressed in detail in the future. 

\section{Mach cones and experimental two-particle correlations}

As discussed above colorless and colored sound can both lead to the occurrence of Mach cone structures in the wake of the away-side leading parton. Those Mach cone structures arre characterized by an opening angle $\phi=\arccos(c_s/v)$.  
In a recent preprint T. Renk and I described a formalism that can be used to track the propagation of such modes through the evolving medium if their dispersion relations are known. I will only shortly outline that mechanism here and refer the reader for further details to \cite{Renk:2005si}. We concentrated on the propagation of  colorless sound. Under the assumption that such a sound wave is created, we track the jet energy loss as a function of spacetime and the resulting Mach cone through the medium evolution. We base those investigations on a space-time evolution model outlined in \cite{Renk:2004gy,Renk:2005hp}. We generate vertices in a Monte Carlo approach with a distribution weighted by the nuclear overlap and determine the jet momentum and parton type by sampling the partonic transverse momentum spectrum as generated by the VNI/BMS parton cascade (see e.g. \cite{Renk:2005yg}). Calculating the radiative energy loss on the near side parton, we decide whether the experimental trigger condition is fulfilled. This restricts the generated vertices to positions close to the surface of the produced matter. Then the direction of the far side parton in the transverse plane is determined taking properly intrinsic $k_T$ into account. With vertex, energy and direction of the away side jet known, we calculate $dE/d\tau$ of the away side parton by using the quenching weights from \cite{Salgado:2003gb}. For illustration of the resulting loss pattern, c.f. Fig. 1 of \cite{Renk:2005si}.
 
\begin{figure}
\begin{minipage}[t]{0.5 \linewidth}
\epsfig{file=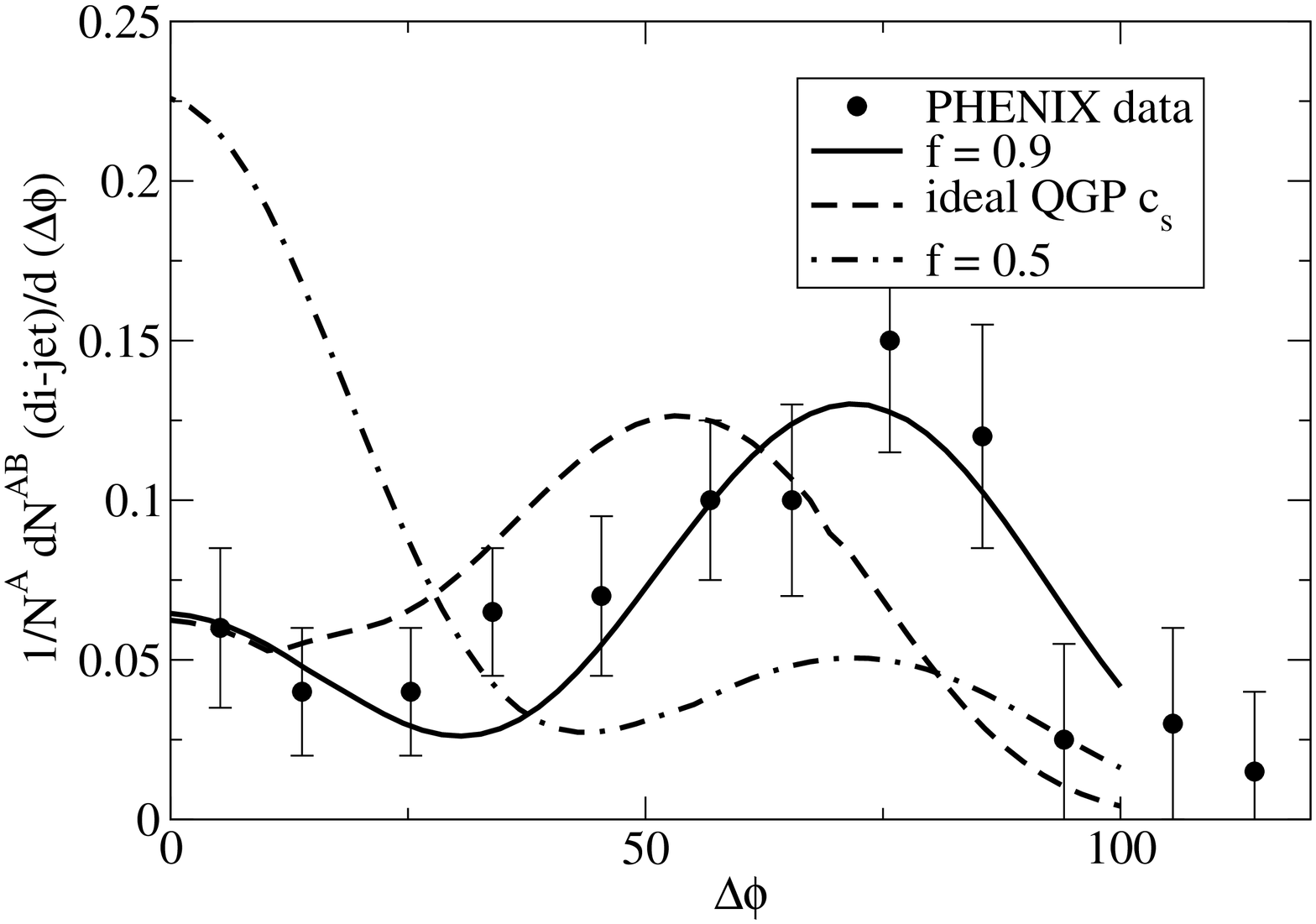, width=7cm, height=7cm}
\end{minipage}
\begin{minipage}[b]{0.5 \linewidth}
\caption{\label{figCORR} {\small This plot shows the two-particle hadronic correlations on the away-side. 0 degrees is chosen as the direction opposite to the near side jet. For details of trigger condition and momentum cuts, see \cite{Adler:2005ee}. All theoretical lines represent calculations using the formalism developed in \cite{Renk:2005si}. 
The parameter $f$ characterizes which fraction of the partonic energy loss is transfered to the sound mode of the medium. The solid line corresponds to $f=0.9$ and a colorless sound determined from lattice EOS. The dashed line illustrates that the sound velocity has to be considerable smaller than the expectation for an ideal QGP during some stage in the evolution.}}
\end{minipage}
\end{figure}

Our further investigation relies on the assumption that a fraction $f$ of the energy lost to the medium excites a collective mode leading to a wave front which propagates with the local speed of sound. Here we specialized on colorless sound with a dispersion relation
$E=c_s p$ which is excited by that energy fraction. The remaining energy  heats the medium and leads to a collective drift along the jet axis conserving longitudinal momentum. $c_s$ is determined locally via the equation of state fitted to describe lattice results. This EOS shows a significant reduction of $c_s$ as one approaches the phase transition but does not lead to a mixed phase. The dispersion relation along with energy momentum deposition determines the initial angle of propagation of the shock front with jet axis (the Mach angle) $\phi=\arccos c_s/v$ and the further propagation of the angle is adjusted with the local speed of sound. The distortion of the cone structure by flow is also taken into account. Two example results for specific away-side partons are shown in Fig. 2 of \cite{Renk:2005si}. Using the Cooper-Frye formula one can determine the resulting momentum spectrum. Focussing on the away-side and employing the appropiate trigger conditions one can compare to PHENIX data for central collisions. Fig. \ref{figCORR} shows the result depending on the sole model parameter, the fraction $f$ of energy transferred to the collective mode. A best fit yields $f\approx0.9$.

Two conclusions
can be drawn immediately. First, a large fraction of the whole energy lost has to excite the shockwave in order to reproduce the data and, second, that the angular position of the maximum at higher degrees is capable of revealing important information about the speed of sound. We observe that on average it has to be significantly {\it smaller} than the ideal QGP value, the data are well compatible with a crossover transition with a soft point in the EOS. 
\section{Conclusions}

To summarize one can conclude from \cite{Renk:2005si}  that a Mach cone mechanism is in good agreement with the two-particle correlation data.

Recently, measurements of 3-particle correlations have been announced by the PHENIX and STAR collaborations. While three particle correlation have to be addressed quantitatively in the framework outlined above in the future, a remark at this point should be appropriated: The Mach cone phenomenon is a collective phenomenon that can - as discussed in \cite{Renk:2005si} - easily involve of the order of hundred particles. While all of them are strongly correlated with the original hard away-side parton (and hence with the observable near side jet) since this correlation is subtracted only longitudinal and transverse momentum balance remain to account for residual correlations inside the cone. With momentum distributed across a significant number of particles, there is no a priori reason to expect strong correlations between two particles.

Many significant physics questions regarding jet energy deposition into the medium have to be addressed in future research: Is there an appropriate  theoretical framework to understand the jet energy deposition mechanism of a fraction of the lost energy? How can the properties of those collective modes indicated by experiment be determined from first principle calculations? Is sound the only collective mode compatible with the data? How can colorful and colorless sound collective modes and their interplay be consistently included? How sensitive is the correlation pattern to the details of the properties of the collective modes?

\vspace{-0.25cm}
\section*{Acknowledgements}\vspace{-0.3cm} {\small  I want to thank  S. A. Bass, D. D. Dietrich, R. Lacey, B. M\"uller, T. Renk,  F. Wang  for helpful discussion and the organizers for their kind invitation.  This work was supported in part by the U.\ S.\ Department of Energy under grant DE-FG02-05ER41367. I thank the A. v. Humboldt Foundation as a Fedor Lynen Fellow and for providing part of the travel funds.}

\end{document}